\documentclass[atmp]{ipart_v1}

\Vol{26}
\Issue{10}
\Year{2022}
\firstpage{3783}

\usepackage[english]{babel}
\usepackage[utf8]{inputenc}

\usepackage{amsthm}
\usepackage{yfonts}

\usepackage{bbm}
\usepackage{bm}
\usepackage{mathrsfs}

\usepackage{amssymb}

\numberwithin{equation}{section}

\theoremstyle{plain}% default

\newcommand{\g}{\gamma}
\newcommand{\ep}{\epsilon}
\newcommand{\m}{\mu}
\newcommand{\f}{\phi}

\renewcommand{\lg}{{\mathfrak{g}}}
\newcommand{\lh}{{\mathfrak{h}}}
\renewcommand{\le}{{\mathfrak{h}}}

\newcommand{\R}{\mathbb{R}}

\newcommand{\trg}{\triangleright}

\newcommand{\Tc}{{\mathcal{T}}}
\renewcommand{\H}{{\mathcal{F}}}

\newcommand{\F}{{\mathrm{F}}}
\newcommand{\G}{{\mathcal{G}}}
\renewcommand{\S}{{\mathcal{S}}}

\newcommand{\raz}{ \rule{0ex}{2.5ex} }
\newcommand{\del}{\partial}
\newcommand{\ds}{\displaystyle}
\newcommand{\pb}[2]{\{\,  {#1} \,,\, {#2} \,\}  }
\newcommand{\bi}[1]{\vec{#1}}
\newcommand{\rmd}{{\mathrm{d}}}

\begin{document}

\title[Hamiltonian analysis of the $BFCG$ theory]{Hamiltonian analysis of the $BFCG$ theory for a strict Lie 2-group}

\author[A. Mikovi\'c, M. \^{A}. Oliveira, and M. Vojinovi\'c]{Aleksandar Mikovi\'c, Miguel \^{A}ngelo Oliveira,\\ and Marko Vojinovi\'c}

\begin{abstract}
We perform a complete Hamiltonian analysis of the $BFCG$ action for a general Lie 2-group by using the Dirac procedure. We show that the resulting dynamical constraints eliminate all local degrees of freedom which implies that the $BFCG$ theory is a topological field theory.
\end{abstract}

\maketitle

\section{\label{intro}Introduction}
Two-groups, or crossed modules, represent a category theory generalization of the concept of a group, see \cite{BL,ML}. Beside their applications in topology, the 2-groups can be used to generalize the notion of a gauge symmetry \cite{BH}. The 2-groups can be also used to formulate state-sum invariants of 4-manifolds \cite{BF}, as well as state-sum models of quantum gravity \cite{MV,M}. In applications of 2-groups to gauge theories and gravity, the central role is played by the $BFCG$ topological field theory, which is a generalization of the usual $BF$ theory such that the Lie group of the $BF$ theory is replaced by a Lie 2-group \cite{GPP,MM}.

In this paper we will carry out the Hamiltonian analysis of the $BFCG$ theory for a general Lie 2-group. This analysis is important because it is the only way to discover what are the physical degrees of freedom (DOF). Also it is the first step towards a canonical quantization of a $BFCG$ theory. Understanding the canonical quantization of a $BFCG$ theory is not only important for formulating new quantum gravity theories, but it will have important implications for constructing new manifold invariants, as well as for the formulation of a Peter-Weyl theorem for a Lie 2-group, see \cite{MM}. Note that the Hamiltonian analysis of a $BFCG$ theory in the special case of the Poincar\'e 2-group has been carried out in \cite{MO,MOV}, while the corresponding canonical quantization has been discussed in \cite{MO}. 

In section \ref{twogroups} we describe 2-groups and the $BFCG$ action. In section \ref{feq} we discuss the field equations and write the $BFCG$ action in terms of the space-time components of the relevant fields. We also present the Bianchi identities which are important for the counting of the local degrees of freedom. In section \ref{short} we perform a gauge-fixed canonical analysis of the $BFCG$ action, which is a generalization of the gauge-fixed canonical analysis introduced in \cite{MO}. In section \ref{full} we perform the complete canonical analysis  by using the Dirac procedure and in section \ref{dof} show that there are no local DOF. In section \ref{conc} we present our conclusions.

\section{\label{twogroups}2-groups and $BFCG$ theory}

A category is an algebraic structure containing objects and maps between the objects, called morphisms, where the morphisms satisfy the composition rules associated to oriented curves (morphisms) joining points  in space (objects), see \cite{BH}. A 2-category contains objects, morphisms and 2-morphisms. The latter are maps between morphisms, and the composition rules are modeled on the composition of oriented curves and the corresponding surfaces, see \cite{BH}. Since a group is a category with one object where all the morphisms are invertible, it is natural to define a 2-group as a 2-category with one object where all the morphism and 2-morphisms are invertible. 

The concrete realization of a 2-group is a crossed module $(G,H,\del,\trg)$, where $G$ and $H$ are groups, $\del: H \to G$ is a homomorphism and $\trg$ is an action of $G$ on $H$ by automorphisms such that
\begin{equation}
\del (g\trg h) = g \del h g^{-1}\,,\qquad \del h \trg h' = h h' h^{-1}\, 
\end{equation}
for all $g\in G$ and $h,h' \in H$. When $G$ and $H$ are Lie groups, one can also construct the differential crossed module $(\lg,\le,\del,\trg)$, where $\lg$ and $\le$ are the corresponding Lie algebras, see \cite{MM}.
   
Given a smooth manifold $M$, a 2-connection can be defined as a pair of forms $(A,\beta)$ such that $A$ is a $\lg$-valued 1-form and $\beta$ is a  $\lh$-valued 2-form. One can associate to $(A,\beta)$ a curvature 2-form $\H$ and a curvature 3-form $\G$ as
\begin{equation}
\H_{(A,\beta)}= d{{A}} + {{A}} \wedge {{A}} - \del\beta  \,,\qquad \G_{({A},\beta)}= d{\beta} + { A}\wedge^\trg \beta \,.\label{2bf}
\end{equation}
Note that $\F_{{A}}= d{{A}} + {{A}} \wedge {{A}}$ is the curvature on the principal bundle $P_G (M)$ and $(A,\beta)$ is a 2-connection on a 2-bundle associated to the 2-group $(G,H)$. 

There are two types of gauge transformations for a 2-connection  $({{A}},\beta)$. Given a smooth map $\f\colon M \to G$, we can define maps
 \begin{equation} {{A}} \mapsto \f^{-1} {{A}}\f+\f^{-1} d \f \,,\qquad \beta\mapsto \f^{-1} \trg \beta \,,\label{tgt}\end{equation}
which will be called a thin gauge transformation. 

Similarly, given a 1-form $\eta$ on $M$ with values in $\lh$, we can define maps
\begin{equation} {{A}} \mapsto {{A}}+\del\eta\,,\qquad\beta\mapsto \beta +d \eta + {{A}} \wedge^\trg \eta+\eta\wedge \eta \, ,\label{fgt}\end{equation}
which will be called a fat gauge transformation. 

The curvature $\F$, the fake curvature $\H$ and the 3-form curvature $\G$ transform under a thin gauge transformation as
\begin{equation}
\F_{{A}} \mapsto \f^{-1} \F_{{A}}\f\, \,,\qquad \H_{({{A}},\beta)} \mapsto \f^{-1} \H_{({{A}},\beta)}\f\,,\qquad \G_{({{A}},\beta)} \mapsto \f^{-1} \trg \G_{({{A}},\beta)} \,,
\end{equation}
while under a fat gauge transformation, they transform as
\begin{equation}
\F_{{A}} \mapsto \F_{{A}} + \del \left(d\eta+ {{A}} \wedge^\trg \eta  + \eta \wedge\eta \right)\,,
\end{equation}
\begin{equation}
\H_{({{A}},\beta)}  \mapsto \H_{({{A}},\beta)} \,,\qquad  
\G_{({A},\beta)} \mapsto \G_{({A},\beta)}+\H_{({{A}},\beta)} \wedge^\trg \eta\,.
\end{equation}

One can construct a topological theory of flat 2-connections by generalizing the $BF$ action to the Lie 2-group case, see \cite{GPP,MM}. In the case when the homomorphism $\del$ is trivial, the corresponding action was constructed in  \cite{GPP}, while the action for the general case was constructed in \cite{MM}. This action is given by
\begin{equation}
  S =   \int_M \langle B \wedge \H_{(A,\beta)}\rangle_\lg +\int_M \langle C \wedge \G_{(A,\beta)} \rangle_\le \,,\label{abfcg}
\end{equation}
where $B$ is a 2-form valued in $\lg$ and $C$ is a 1-form valued in $\lh$. The evaluations
$\langle,\rangle_\lg$ and $\langle,\rangle_\le$ are $G$-invariant, {bilinear}, non-degenerate and symmetric forms in the corresponding Lie algebras.

The $BFCG$ action (\ref{abfcg}) will be invariant under a thin gauge transformation if
\begin{equation} C \to \f^{-1}\trg C \,,\qquad B \to \f^{-1} B \f \,,\label{gbct}\end{equation}
while the invariance under a fat gauge transformation is achieved if the fields $B$ and $C$ transform as
\begin{equation}B \mapsto B+C\wedge^{\Tc} \eta \,,\qquad C \mapsto C   \,.\label{bcf}\end{equation}
The antisymmetric map $\Tc\colon \le \times \le \to \lg$ is defined as
\begin{equation}\label{tau}
\langle  \Tc(u,v), Z \rangle_\lg=-\langle  u, Z \trg  v\rangle_\le\,, \qquad u,v \in \le \,,\, Z \in \lg \,.
\end{equation}
Also note that $C\wedge^{\Tc} \eta$ is the antisymmetrization of $\Tc(C,\eta)$, see \cite{MM} for details.

\section{\label{feq}Spacetime components of the action}

In order to perform the canonical analysis of the $BFCG$ action (\ref{abfcg}), we need to write it in terms of the spacetime components of the relevant fields.

Let $T_a$ be a basis in $\lg$ and $\tau_\alpha$ a basis in $\le$. The structure constants are defined by 
\begin{equation}
[T_a , T_b ] = f^c{}_{ab} \,T_c\,,\qquad [\tau_\alpha , \tau_\beta ] = \f^{\gamma}{}_{\alpha\beta} \,\tau_\gamma \,.
\end{equation}    
The homomorphisms $\del$ and $\trg$ then act as
\begin{equation}
\del \tau_\alpha = \del_\alpha{}^a \,T_a \,,\qquad T_a \trg \tau_\alpha = \trg^{\beta}{}_{a\alpha} \, \tau_\beta
\end{equation}
and satisfy the following relations, 
\begin{equation} 
\trg^{\beta}{}_{a\alpha}{} \,\del{_\beta{}^b} = \del{_\alpha{}^c} \,f^b{}_{ac} \,,
\qquad 
\del{_\alpha{}^a} \,\trg^{\gamma}{}_{a\beta} = \f^{\gamma}{}_{\alpha\beta} \,.
\end{equation} 
Also, the following relation 
\begin{equation}  
f^a{}_{bc} \trg_{\alpha a \beta} = \trg_{\alpha [b| \g}\trg^\g{}_{|c] \beta}\,, 
\end{equation} 
will be useful, where $X_{[bc]} = X_{bc} - X_{cb} $.

The structure constants satisfy the Jacobi identities
\begin{equation}
f^d{}_{ac} \,f^c{}_{be} = f^c{}_{a [b|} \,f^d{}_{c|e]} \,, \qquad \f^{\gamma}{}_{\alpha\delta} \,\f^{\delta}{}_{\beta\epsilon} = \f^{\delta}{}_{\alpha [\beta|} \,\f^{\gamma}{}_{\delta|\epsilon]} \,.\label{jac}
\end{equation}
We have also 
\begin{equation}
\langle X\wedge Y \rangle_\lg = X^a \wedge Y^b \langle T_a , T_b \rangle_\lg = X^a \wedge Y^b Q_{ab}\,,
\end{equation}
and
\begin{equation}
\langle U\wedge V \rangle_\lh = U^{\alpha} \wedge V^{\beta} \langle \tau_{\alpha} , \tau_{\beta} \rangle_\lh = U^{\alpha} \wedge V^{\beta} q_{\alpha\beta}\,.
\end{equation}
The fake curvature can be written as 
\begin{equation}
 \H_{(A,\beta)}= \frac{1}{2} \H{^b{}_{\mu\nu}} T_{b}\,dx^\mu\wedge dx^\nu
\end{equation}
where 
\begin{equation}\label{fakecurvature}
\H{^b{}_{\mu\nu}}=\left.\partial_\mu A{^b{}_\nu}-\partial_\nu A{^b{}_\mu}+f{^b{}_{cd}}{A{^c{}_\mu}}{A{^d{}_\nu}}- \del{_\alpha{}^b} \beta{^\alpha{}_{\mu\nu}}\right. \,.
\end{equation}
The curvature 3-form can be written as
\begin{equation}
 \G_{(A,\beta)} = \ds\frac{1}{6}\G{^\alpha{}_{\mu\nu\rho}}\tau_\alpha dx^\mu\wedge dx^\nu \wedge dx^\rho
\end{equation}
where
\begin{equation}\label{2curvature}
\G{^\alpha{}_{\mu\nu\rho}}= \del_{ [\mu}\beta{^\alpha{}_{\nu\rho]}} + A{^a{}_{[\mu}} \beta{^\gamma{}_{\nu\rho]}} \trg{^\alpha{}_{a\gamma}}\,.
\end{equation}
$X_{[\mu\nu\rho]}$
denotes a total antisymmetrization of indices, given by
\begin{equation}
\sum_{p\in S_3} (-1)^p \, X_{p(\mu\nu\rho)} \,.
\end{equation} 
where $p$ is a permutation and $(-1)^p$ is the parity of $p$.
  
The $BFCG$ action then becomes
\begin{equation}\label{action}
S=\int_M \rmd^4 x \,\epsilon^{\mu\nu\rho\sigma}\left(\frac{1}{4}\,B{^a{}_{\mu\nu}}\,\H{^b{}_{\rho\sigma}}\, Q_{ab}+\frac{1}{6}\, C{^\alpha{}_\mu}\,
\G{^\beta{}_{\nu\rho\sigma}}\, q_{\alpha\beta}\right)\,.
\end{equation}
To simplify notation we use $Q$ and $q$ to lower $\lg$ and $\le$ Lie algebra indices respectively, for example,
\begin{equation}
B_b=B^a Q_{ab} \qquad \beta_\beta = \beta^\alpha q_{\alpha\beta}\,.
\end{equation}
We also use the same symbol for $\del{_\alpha{}^a}$ and $\del{^\alpha{}_a}=q^{\alpha\beta}\del{_\beta{}^b}Q_{ba}$ and $\trg{_{\alpha a\gamma}}=\trg{^\beta{}_{a\gamma}}q_{\beta\alpha}$. This last quantity (that was in fact already defined in (\ref{tau})) is antisymmetric in $\alpha \gamma$, that is  $\trg{_{\alpha a\gamma}}=-\trg{_{\gamma a\alpha}}$, and we have as a consequence
\begin{equation}
\begin{aligned}
C^\alpha \nabla^\trg_\mu \beta_\alpha & =  \ds C{^\alpha}\left(\del_\mu \beta{_{\alpha}}+A{^a{}_\mu}\trg{_{\alpha a\gamma}}\beta{^\gamma}\right) \vphantom{\int} \\
 & =  \ds -\left(\del_\mu C_\alpha +A{^a{}_\mu}\trg{_{\alpha a\gamma}}C^\gamma\right)\beta{^\alpha}+\del_\mu\left(C^\alpha \beta_\alpha\right) \vphantom{\int} \\
 & =  \ds -\nabla^\trg_\mu\left( C^\alpha \right)\beta_\alpha + \del_\mu\left(C^\alpha \beta_\alpha\right)\,, \vphantom{\int} \\
\end{aligned}
\end{equation}
where $\nabla^\trg_\mu$ is defined as the quantity in parenthesis.

The equations of motion are obtained by equating to zero the variational derivatives of the action with respect to all fields. The variational derivatives with respect to $B$ and $C$  give
\begin{equation}\label{F1}
\H{^b{}_{\mu\nu}}=0\,,  \qquad   \G{^\alpha{}_{\mu\nu\rho}}=0 \,, 
\end{equation}
while the variational derivatives with respect to $A$ and $\beta$ give 
\begin{equation}\label{F2}
\begin{aligned}
\ds\epsilon^{\mu\nu\rho\sigma}\left(\nabla_\mu B{_{a \nu\rho}}+\beta{^\alpha{}_{\mu\nu}}\trg{_{\alpha a\beta}}C{^\beta{}_\rho}\vphantom{\ds\frac{1}{2}}\right) & =  0\,, \\ 
\ds\epsilon^{\mu\nu\rho\sigma}\left( \nabla^\trg_\mu C{^\alpha{}_\nu}-\ds\frac{1}{2}\del^\alpha{}_a B{^a{}_{\mu\nu}}   \right) & =  0 \,. \\
\end{aligned}
\end{equation}
We will also use the Bianchi identities (BI) associated to the 1-form fields $A$ and $C$. Namely, the corresponding 2-form curvatures
\begin{equation}
 F^a = dA^a + f^a{}_{bc}\,A^b \wedge A^c \,,\qquad T^\alpha = dC^\alpha + \trg^\alpha{}_{a\beta} \, A^a \wedge C^\beta \,,
\end{equation}
satisfy the following Bianchi identities  
\begin{equation} 
\ep^{\lambda\m\nu\rho} \,\nabla_\m F^a{}_{\nu\rho} = 0 \,,\label{bia}
\end{equation}
and
\begin{equation}
\ep^{\lambda\m\nu\rho}\left( \nabla^\trg_\m T^\alpha{}_{\nu\rho} -\,\trg^\alpha{}_{a\beta} F^a{}_{\m\nu}C^\beta{}_\rho \right) = 0 \,.\label{bic}
\end{equation}
There are also the BI associated with the 2-form fields $B$ and $\beta$. The corresponding 3-form curvatures are given by
\begin{equation}
 G^a = d B^a + f^a{}_{bc}\,A^b \wedge B^c \,, \qquad \G^\alpha = d\beta^\alpha + \trg^\alpha{}_{a\g} \,A^a \wedge \beta^\g \,,
\end{equation}
so that
\begin{equation}
 \ep^{\lambda\m\nu\rho}\left( \frac{2}{3} \nabla_\lambda \,G^a{}_{\m\nu\rho} -f^a{}_{bc} F^b{}_{\lambda\m}\,B^c{}_{\nu\rho}\right) = 0 \label{b2c}
 \end{equation}
and
\begin{equation} 
\ep^{\lambda\m\nu\rho}\left( \frac{2}{3} \nabla^\trg_\lambda \,\G^\alpha{}_{\m\nu\rho} - \trg^\alpha{}_{a\g} \,F^a{}_{\lambda\m}\,\beta^\g{}_{\nu\rho}\right) = 0 \,.\label{b2d}
\end{equation}
   
\section{\label{short}A gauge-fixed canonical analysis}

We will assume that $M=\Sigma\times {\R}$ and that $t$ is a coordinate on $\R$ while $\{ x^i | i=1,2,3\}$ is a local coordinate chart on $\Sigma$. We can split the $BFCG$ fields into temporal and spatial components by using
\begin{equation}
x^\mu =(x^0,x^i)=(t,\vec x)
\end{equation}
and  $U_\mu = (U_0 , U_i)$. For example 
\begin{equation}
\del_{\mu} U_{\nu} = (\partial_0 U_0 , \partial_0 U_i , \partial_i U_0 , \partial_i U_j) 
\end{equation}
and
\begin{equation}
\begin{aligned}
  \epsilon^{\mu\nu\rho\sigma} \partial_\mu U_{\nu} \partial_{\rho}V_ \sigma & =  \epsilon^{0ijk} \partial_0 U_{i} \partial_{j}V_ k + \epsilon^{i0jk} \partial_i U_{0} \partial_j V_k \vphantom{\ds\int}\\
  &\quad  + \epsilon^{ij0k} \partial_i U_{j} \partial_0 V_k + \epsilon^{ijk0} \partial_i U_{j} \partial_k V_0 \vphantom{\ds\int} \\
& =  \epsilon^{ijk}\left( \dot{U}_{i} \partial_j V_k - \partial_i U_{0} \partial_j V_k + \partial_i U_{j} \dot{V_k} -  \partial_i U_{j} \partial_k V_0  \right) \,, \vphantom{\ds\int} \\
\end{aligned}
\end{equation}
where $\dot{X} = \partial_0 X$ and throughout the rest of the paper, $\epsilon^{ijk} \equiv \epsilon^{0ijk}$.

The $BFCG$ action can be then written as 
\begin{equation}
 S = \int_{t_1}^{t_2} dt \, L (t) \,, 
\end{equation}
where the Lagrangian $L$ is given by 
\begin{equation}
L=\int_\Sigma \rmd^3\bi{x} \left[\pi(A){_a{}^i}\dot{A}{^a{}_i}+\frac{1}{2}\pi(\beta){_\alpha{}^{ij}}
\dot{\beta}{^\alpha{}_{ij}}\right]- H 
\end{equation}
and
\begin{equation}
\begin{aligned}
  H & =  \ds -\int_\Sigma\rmd^3\bi{x}\left[\frac{1}{2}\epsilon^{ijk}B_{a\,0i}\S(\H){^a{}_{jk}}+C_{\alpha 0}\S(\G)^\alpha + A{^a{}_0}\S(BC\beta)_a \right. \\
    &\qquad \ds \left. +\beta{^\alpha{}_{k0}}\S(CA){_\alpha{}^k} + \del_i\left(\pi(A){_a{}^i}A{^a{}_0}-\pi(\beta){_\alpha{}^{ij}}\beta{^\alpha{}_{j0}}\right) \vphantom{\ds\frac{1}{2}} \right]\,.
\end{aligned}
\end{equation}
The fields $\pi (A)$ and $\pi(\beta)$ are given by
\begin{equation}\label{pis}
	\pi(A){_a{}^i}=\frac{1}{2}\,\epsilon^{ijk}B{_{a\,jk}}\,,\qquad\qquad \pi(\beta){_\alpha{}^{ij}}=-\,\epsilon^{ijk}C_{\alpha\,k}\,,
\end{equation}
while
\begin{equation}\label{cnstr}
\begin{aligned}
\S(\H){^a{}_{ij}}&\equiv\H{^a{}_{ij}}\,,\\
\S(\G)^\alpha&\equiv \ds \frac{1}{6}\epsilon^{ijk}\G{^\alpha{} _{ijk}}\,,\\
\S(BC\beta)_a&\equiv\nabla_k\pi(A){_a{}^k}-\ds\frac{1}{2}\pi(\beta){_\alpha{}^{jk}}\trg{^\alpha{}_{a\beta}}\beta{^\beta{}_{jk}}\,,\\
\S(CB){_\alpha{}^k}&\equiv\ds \frac{1}{2}\nabla^\trg_j\pi(\beta){_\alpha{}^{jk}}+\del_\alpha{}^a \pi(A){_a{}^k}\,.
\end{aligned}
\end{equation}

From these equations we see that the $BFCG$ Lagrangian has the form
\begin{equation}
L =\sum_m P_m \dot{Q}_m - H = \sum_m P_m \dot{Q}_m - \sum_n \lambda_n G_n (P,Q) \,.
\end{equation}
According to the theorem proved in \cite{MO}, such a Lagrangian is a result of the Dirac procedure in the gauge $P(\lambda_n ) =0$ if the constraints $G_n(P,Q)$ are of the first class with respect to the $(P,Q)$ Poisson bracket, i.e. form a closed algebra under the Poison bracket defined by 
\begin{equation}\{A ,B \}  = \sum_n \left(\frac{\partial A}{\partial Q_n} \frac{\partial B}{\partial P_n} - 
\frac{\partial A}{\partial P_n} \frac{\partial B}{\partial Q_n}\right) \,.\label{gfpb}\end{equation}

It is straightforward to verify that the constraints from (\ref{cnstr}) are of the first class, by using the PB (\ref{gfpb}). The non-zero PB are then given by
\begin{equation}
\begin{aligned}
\pb{A^{a}{}_{i}(x)}{\pi(A)_{b}{}^{j}(y)} &\raz  =   \delta^a_{b} \delta_{i}^{j} \delta^{(3)}(x-y)\,, \\
\pb{\beta^{\alpha}{}_{ij}(x)}{\pi(\beta)_{\beta}{}^{kl}(y)}  &\raz  =    \delta^\alpha_{\beta} \delta^{k}_{[ i} \delta^{l}_{j]} \delta^{(3)}(x-y)\,, 
\end{aligned}
\end{equation}
where $x=\vec x$, $y=\vec y$ and $\delta^{(3)}(x-y)$ is the three-dimensional Dirac delta function.

The Poisson-bracket algebra for the constraints from (\ref{cnstr}) is then given by
\begin{equation}\label{algS}
\begin{aligned}
\pb{\S(\H){^a{}_{ij}}(x)}{\S(BC\beta)_b (y)} & =  2f{^a{}_{bc}}\S(\H){^c{}_{ij}(x)}\,\delta^{(3)}(x-y)\,,\\
\pb{\S(\G)^\alpha(x)}{\S(CB){_\beta{}^k}(y)} & =  \epsilon^{ijk}\trg{^\alpha{}_{c \beta}} \S(\H){^c{}_{ij}}(x)\,\delta^{(3)}(x-y)\,,\\
\pb{\S(BC\beta)_a (x)}{\S(BC\beta)_b (y)} & = 2 f{^c{}_{ab}}     \S(BC\beta)_c (x) \,\delta^{(3)}(x-y)\,,\\
\pb{\S(\G)^\alpha(x)}{\S(BC\beta)_a(y)}& = 2\trg{^\alpha{}_{a\beta}}S(\G)^\beta(x)\,\delta^{(3)}(x-y)\,,\\
\pb{\S(CB){_\alpha{}^k}(x)}{\S(BC\beta)_a(y)} & =  \trg{^\beta{}_{a\alpha}}\S(BC){_\beta{}^k}(x)\,\delta^{(3)}(x-y)\,,
\end{aligned}
\end{equation}
which confirms that they are of the first class. Hence the constraints of the action (\ref{action}) correspond to the constraints of the Dirac analysis in the gauge 
\begin{equation}
\pi(B^a{}_{0i})=\pi(C^\alpha{}_0)=\pi(A^a{}_0)=\pi(\beta^\alpha{}_{0i})=0\,.
\end{equation}

\section{\label{full}The complete canonical analysis}

The analysis in the previous section has an implicit gauge fixing. To see this, we can perform the complete canonical analysis by using the Dirac procedure, see \cite{B}. For this we consider the Lagrangian
\begin{equation}\label{lagrangean}
L=\int_\Sigma \rmd^3\bi{x}\, \epsilon^{\mu\nu\rho\sigma}\left(\frac{1}{4}\,B{^a{}_{\mu\nu}}\,\H{^b{}_{\rho\sigma}}\,Q_{ab}+\frac{1}{6}\, C{^\alpha{}_\mu}\,
\G{^\beta{}_{\nu\rho\sigma}}\, q_{\alpha\beta}\right)\,,
\end{equation}
and calculate the momenta (the functional derivatives of the Lagrangian with respect to the time derivatives of the variables) for all variables $B{^a{}_{\mu\nu}}$, $A{^a{}_\mu}$, $C{^\alpha{}_\mu}$ and $\beta{^\alpha{}_{\mu\nu}}$,
\begin{equation}
\begin{aligned}
\pi(B){_{a}{}^{\mu\nu}} &\raz  =   \ds \frac{\delta L}{\delta \del_0 B{^{a}{}_{\mu\nu}} } =  0\,, \\
\pi(C){_\alpha{}^\mu} & \raz =  \ds \frac{\delta L}{\delta \del_0 C{^\alpha{}_\mu}}  =  0 \,, \vphantom{\ds\int^A} \\
\pi(A){_{a}{}^{\mu}} & \raz =  \ds \frac{\delta L}{\delta \del_0 A{^{a}{}_{\mu}}}  =  \ds \frac{1}{2}\epsilon^{0\mu\nu\rho} B_{a\nu\rho}\,, \vphantom{\ds\int^A} \\
\pi(\beta){_\alpha{}^{\mu\nu}} & \raz =   \ds \frac{\delta L}{\delta \del_0 \beta{^\alpha{}_{\mu\nu}}}  =  - \epsilon^{0\mu\nu\rho} C_{\alpha\rho}\,. \vphantom{\ds\int^A} \\
\end{aligned}
\end{equation}
All of these momenta give rise to primary constraints since none of them can be inverted for the  time derivatives of the variables, 
\begin{equation}
\begin{aligned}
P(B){_{a}{}^{\mu\nu} }& \raz\equiv  \pi(B){_{a}{}^{\mu\nu}} \approx 0\,, \\
P(C){_\alpha{}^{\mu}} & \raz\equiv  \ds \pi(C){_\alpha{}^{\mu}}  \approx 0\,, \\
P(A){_{a}{}^{\mu}} & \raz\equiv  \pi(A){_{a}{}^{\mu}} -\ds  \frac{1}{2}\epsilon^{0\mu\nu\rho} B_{a\nu\rho} \approx 0\,, \\
P(\beta){_\alpha{}^{\mu\nu}} & \raz\equiv  \pi(\beta){_\alpha{}^{\mu\nu}} + \epsilon^{0\mu\nu\rho}C_{\alpha\rho} \approx 0\,. 
\end{aligned}
\end{equation}

We use the weak equality ``$\approx$'' for the equality that holds on a subspace of the phase space, while the equality that holds for any point of the phase space will be called ``strong'' and it is denoted by the usual symbol ``$=$''. We will also use the expressions ``on-shell'' and ``off-shell'' for strong and weak equalities, respectively.

We will use the following fundamental Poisson brackets 
\begin{equation}
\begin{aligned}
\pb{B{^{a}{}_{\mu\nu}}(x)}{\pi(B){_{b}{}^{\rho\sigma}(y)}} & \raz =   \delta^a_{b}  \delta^{\rho}_{[ \mu} \delta^{\sigma}_{\nu]} \,\delta^{(3)}(x-y)\,, \\
\pb{C{^\alpha{}_{\mu}}(x)}{\pi(C){_\beta{}^{\nu}}(y)} & \raz =  \delta^\alpha_\beta \delta^{\nu}_{\mu} \,\delta^{(3)}(x-y)\,, \\
\pb{A{^{a}{}_{\mu}}(x)}{\pi(A){_{b}{}^{\nu}}(y)} & \raz =  \delta^a_{b}  \delta^{\nu}_{\mu} \,\delta^{(3)}(x-y)\,, \\
\pb{\beta{^\alpha{}_{\mu\nu}}(x)}{\pi(\beta){_\beta{}^{\rho\sigma}}(y)} & \raz =  \delta^\alpha_\beta \,\delta^{\rho}_{[\mu} \delta^{\sigma}_{\nu]} \,\delta^{(3)}(x-y)\,, \\
\end{aligned}
\end{equation}
to calculate the algebra between the primary constraints. We obtain 
\begin{equation} \label{AlgebraPrimarnihVeza}
\begin{aligned}
\pb{P(B){_a{}^{jk}}(x)}{P(A){_{b}{}^i}(y)} & =  \epsilon^{0ijk}\,Q_{ab}(x) \,\delta^{(3)}(x-y)\,, \\
\pb{P(C){_\alpha{}^{k}}(x)}{P(\beta){_b{}^{ij}}(y)} & =  - \epsilon^{0ijk}\, q_{ab}(x)\, \delta^{(3)}(x-y)\,,
\end{aligned}
\end{equation}
while all other Poisson brackets vanish. 

The canonical on-shell Hamiltonian is defined by
\begin{equation}\label{Hcan}
\begin{aligned}
  H_c & =  \ds \int_{\Sigma} \rmd^3\bi{x} \left[ \frac{1}{2} \pi(B){_{a}{}^{\mu\nu}} \del_0 B{^{a}{}_{\mu\nu}} + \pi(C){_\alpha{}^{\mu}} \del_0 C{^\alpha{}_{\mu}} \right. \\
    &  \hphantom{mmmmm} \ds \left. + \pi(A){_{a}{}^{\mu}} \del_0 A{^{a}{}_{\mu}} + \frac{1}{2} \pi(\beta){_\alpha{}^{\mu\nu}} \del_0 \beta{^\alpha{}_{\mu\nu}} \right] -L \,. \\
\end{aligned}
\end{equation}
By using (\ref{lagrangean}), (\ref{fakecurvature}) and (\ref{2curvature}) we can rewrite the Hamiltonian (\ref{Hcan}) such that all the velocities are multiplied by the first class constraints. Therefore in an on-shell quantity they drop out, so that
\begin{equation}
\begin{aligned}
  H_c & =  \ds - \int \rmd^3\bi{x}\, \epsilon^{0ijk} \left[ \frac{1}{2} B_{a0i} \H{^{a}{}_{jk}} + \frac{1}{6} C_{\alpha0} \G{^\alpha{}_{ijk}} \right. \\
    &  \hphantom{mmmmmmmm}\ds + \beta{^\alpha{}_{0k}} \left(\nabla^\trg_iC_{\alpha j}-\frac{1}{2}\del{_\alpha{}^a}B_{a\,{ij}}\right) \\
    &  \hphantom{mmmmmmmm}\ds \left.+ \frac{1}{2} A{^a{}_0}\left( \nabla_i B{_{a\,jk}} - C{^\alpha{}_i}\trg_{\alpha a \beta} \beta{^\beta{}_{jk}} \right) \right] \,. \\
\end{aligned}
\end{equation}
This expression does not depend on any of the canonical momenta and it contains only the fields and their spatial derivatives. By adding a Lagrange multiplier $\lambda$ for each of the primary constraints we can build the off-shell Hamiltonian, which is given by

\begin{equation} \label{TotalniHamiltonijan}
\begin{aligned}
  H_T & =  \ds H_c + \int \rmd^3\bi{x} \left[ \lambda(C){^\alpha{}_{\mu}} P(C){_\alpha{}^{\mu}} + \lambda(A){^{a}{}_{\mu}} P(A){_{a}{}^{\mu}} \vphantom{\ds\frac{1}{2}} \right. \\
    &  \hphantom{mmmmmmmm} \ds \left. + \frac{1}{2} \lambda(B){^{a}{}_{\mu\nu}} P(B){_{a}{}^{\mu\nu}} + \frac{1}{2} \lambda(\beta){^\alpha{}_{\mu\nu}} P(\beta){_a{}^{\mu\nu}} \right] \,. \\
\end{aligned}
\end{equation}

Since the primary constraints must be preserved in time, we must impose the following requirement
\begin{equation}
\dot{P} \equiv \pb{P}{H_T} \approx 0\,,\label{pcc}
\end{equation}
for each primary constraint $P$. By using the consistency condition (\ref{pcc}) for the primary constraints ${P}(B)_{a}{}^{0i}$, ${P}(C){_\alpha{}^0}$, ${P}(\beta){_\alpha{}^{0i}}$ and ${P}(A){_{a}{}^0}$
we obtain the secondary constraints $\S$
\begin{equation} \label{SekundarneVeze}
\begin{aligned}
\S(\H)^{a}{}_{jk} & \equiv  \H{^{a}{}_{jk}} \approx 0, \\
\S(\G)^\alpha & \equiv \ds \frac{1}{6}\epsilon^{0ijk}\G{^\alpha{}_{ijk}} \approx 0, \\
\S(CB)_{\alpha ij} & \equiv  \left(\nabla^\trg_{[ i|}C_{\alpha |j]}-\del{_\alpha{}^a}B_{a\,{ij}}\right) \approx 0, \\
\S(BC\beta)_{a} & \equiv  \ds\frac{1}{2}\epsilon^{0ijk} \left( \nabla_i B{_{a\,jk}} -\ds C{^\alpha{}_i}\trg_{\alpha a \beta} \beta{^\beta{}_{jk}}  \right) \approx 0 \,, \\
\end{aligned}
\end{equation}
while in the case of ${P}(B){_{a}{}^{jk}}$, ${P}(C){_\alpha{}^k}$, ${P}(\beta){_\alpha{}^{jk}}$ and ${P}(A){_{a}{}^k}$
the consistency condition determines the following Lagrange multipliers
\begin{equation}\label{lagrangem}
\begin{aligned}
\lambda(A){^{a}{}_i} & \approx  \ds \nabla_i A{^{a}{}_0}-\del{_\alpha{}^a}\beta{^\alpha{}_{i0}}\,, \vphantom{\frac{1}{2}} \\
\lambda(\beta){^\alpha{}_{ij}} & \approx  \ds  \nabla^\trg_{[ i|} \beta{^\alpha{}_{0 |j]}} -  \beta{^\beta{}_{ij}}\trg^\alpha{}_{a\beta} A{^a{}_0} \,, \vphantom{\frac{1}{2}} \\
\lambda(C){^\alpha{}_i} & \approx  \ds \nabla^\trg_i C{^\alpha{}_0} + C{^\beta{}_i} \trg_{\beta a}{}^{\alpha} A{^a{}_{0}}  \,, \vphantom{\frac{1}{2}} \\
\lambda(B){^{a}{}_{ij}} & \approx  \ds  \nabla_{[ i|} B{^{a}{}_{0 |j]}} -  C{^\alpha{}_0}\trg_\alpha{}^a{}_\g \beta{^\g{}_{ij}}  \vphantom{\frac{1}{2}} \\ 
 & \quad  + f^a{}_{bc}A{^b{}_0}B{^c{}_{ij}} - C{^{\alpha}{}_{[ i|}}\trg_\alpha{}^a{}_\g \beta{^{\g}{}_{0 |j] }}  \,. \vphantom{\frac{1}{2}} \\
\end{aligned}
\end{equation}
The consistency conditions of the secondary constraints (\ref{SekundarneVeze}) turn out to be identically satisfied, and produce no new constraints. Note that the consistency conditions leave the Lagrange multipliers
\begin{equation}
\lambda(A){^{a}{}_0}\,, \qquad
\lambda(\beta){^\alpha{}_{0i}}\,, \qquad
\lambda(C){^\alpha{}_0}\,, \qquad
\lambda(B){^{a}{}_{0i}}\,,
\end{equation}
undetermined.

By using (\ref{lagrangem}), the total Hamiltonian can be written as
\begin{equation} \label{TotHamiltonijanKombVezaPrveKlase}
\begin{aligned}
H_T & =  \ds \int_\Sigma d^3{x} \left[  \lambda(B){^{a}{}_{0i}} \,\phi(B){_{a}{}^i} + \lambda(C){^\alpha{}_0} \,\phi(C)_\alpha   \vphantom{\frac{1}{2}} \right. \\
  &  \hphantom{mmmmm} \ds + \lambda(\beta){^\alpha{}_i} \,\phi(\beta){_\alpha{}^i} +  \lambda(A)^{a} \,\phi(A)_{a} - B_{a0i} \,\phi(\H)^{ai}  \\
  &  \hphantom{mmmmm} \ds \left. - C_{\alpha0} \,\phi(\G)^\alpha - \beta_{\alpha0i} \,\phi(CB)^{\alpha i} - A_{a0} \,\phi(BC\beta)^{a} \vphantom{\frac{1}{2}} \right]\,, \\
\end{aligned}
\end{equation}
where  
\begin{equation}\label{phi}
\begin{aligned}
\phi(B){_{a}{}^i} & =  \ds P(B){_{a}{}^{0i}}\,, \vphantom{\ds\int} \\
\phi(C)_\alpha & =  \ds P(C){_\alpha{}^0}\,, \vphantom{\ds\int} \\
\phi(\beta){_\alpha{}^i} & =  \ds P(\beta){_\alpha{}^{0i}}\,, \vphantom{\ds\int} \\
\phi(A)_{ab} & =  \ds P(A){_{a}{}^0}\,, \vphantom{\ds\int} \\
\phi(\H)^{ai} & =  \ds \frac{1}{2}\epsilon^{0ijk} \S(\H){^{a}{}_{jk}} - \nabla_j P(B)^{aij}\,, \vphantom{\ds\int} \\
\phi(\G)^\alpha & =  \ds  \S(\G)^\alpha + \nabla^\trg_i P(C)^{\alpha i} - \frac{1}{2} \beta_{\beta ij}\trg^\beta{}_a{}^\alpha P(B)^{a ij}\,, \vphantom{\ds\int} \\
\phi(CB)^{\alpha i} & =  \ds \frac{1}{2} \epsilon^{0ijk} \S(CB){^\alpha{}_{jk}} - \nabla^\trg_j P(\beta)^{a ij} \vphantom{\ds\int} \\
 & \quad \ds - C_{\beta j} \trg^\beta{}_a{}^\alpha P(B)^{a ij}+ \del{^\alpha{}_a}P(A){^{a i}}, \vphantom{\ds\int} \\
\phi(BC\beta)^{a} & =  \ds \S(BC\beta)^{a} + \nabla_i P(A)^{ai} -\frac{1}{2} f{^a{}_{bc}} B{^{b}{}_{ij}} P(B)^{cij} \vphantom{\ds\int} \\
 & \quad \ds -  C{^{\alpha}{}_i}\trg_\alpha{}^a{}_\beta P(e)^{\beta i} - \frac{1}{2}\beta{^{\alpha}{}_{ij}} \trg_\alpha{}^a{}_\beta P(\beta){^{\beta ij}} \,,  \\
\end{aligned}
\end{equation}
are the first-class constraints, while
\begin{equation}\label{chi}
\begin{array}{lll}
\chi(B){_{a}{}^{jk}}=P(B){_{a}{}^{jk}}\,, & & \chi(C){_\alpha{}^i}=P(C){_\alpha{}^i}\,, \vphantom{\ds\int} \\
\chi(A){_{a}{}^i}=P(A){_{a}{}^i}\,, & & \chi(\beta){_\alpha{}^{ij}}=P(\beta){_\alpha{}^{ij}}\,. \vphantom{\ds\int} 
\end{array}
\end{equation}\label{phiA}
are the second-class constraints. 

The PB algebra of the first-class constraints is given by
\begin{equation} \label{AVPK}
\begin{aligned}
\pb{\phi(\G)^\alpha (x)}{\phi(CB)^{\beta i}(y)} & =  \ds \trg{^\alpha{}_{a \beta}} \,\phi(\H)^{ai}(x)\, \delta^{(3)}(x-y) \,,  \\
\pb{\phi(\G)^\alpha (x)}{\phi(BC\beta)_{a}(y)} & =  \ds 2\trg{^\alpha{}_{a\beta}} \, \phi(\G)^{\beta}(x)\, \delta^{(3)}(x-y) \,, \vphantom{\ds\int} \\
\pb{\phi(BC){_\alpha{}^k}(x)}{\phi(BC\beta)_a(y)} & =  \trg{^\beta{}_{a\alpha}}\,\phi(BC){_\beta{}^k}(x)\,\delta^{(3)}(x-y)\,,\\
\pb{\phi(\H){^a{}_{ij}}(x)}{\phi(BC\beta)_b(y)} & =  2f{^a{}_{bc}}\,\phi(\H){^c{}_{ij}}(x)\,\delta^{(3)}(x-y)\,, \\
\pb{\phi(BC\beta)_a(x)}{\phi(BC\beta)_b(y)} & = 2 f{^c{}_{ab}}  \,   \phi(BC\beta)_c (x)\,\delta^{(3)}(x-y)\,.
\end{aligned}
\end{equation}
The PB algebra between the first and the second-class constraints is given by
 \begin{equation} \label{AlgebraVezaMesoviteKlase}
\begin{aligned}
\pb{\phi(\H)^{ai}(x)}{\chi(A)_{b}{}^j(y)} & =  -f{^a{}_{bc}} \,\chi(B)^{c}{}{}^{ij}(x)\, \delta^{(3)}(x-y) \,, \vphantom{\ds\int} \\
\pb{\phi(\G)^\alpha (x)}{\chi(A)_{a}{}^i(y)} & =  -\trg{^\alpha{}_{a\g}} \,\chi(C)^{\g}{}^i (x)\,\delta^{(3)}(x-y) \,, \vphantom{\ds\int} \\
\pb{\phi(\G)^\alpha(x)}{\chi(\beta)_\beta{}^{ij}(y)} & =  \trg{^\alpha{}_{a\beta}}\, \chi(B)^{\g}{}{}^{ij}(x)\, \delta^{(3)} (x-y)\,, \vphantom{\ds\int} \\
\pb{\phi(CB)^{\alpha i}(x)}{\chi(A)_{a}{}^j(y)} & =  -\trg{^\alpha{}_{a}{}^\g}\, \chi(\beta)_{\g}{}^{ij}(x)\, \delta^{(3)}(x-y) \,, \vphantom{\ds\int} \\
\pb{\phi(CB)^{\alpha i}(x)}{\chi(C)_\beta{}^j(y)} & =  \trg{^{\alpha}{}_{a \beta}}\,\chi(B)^a{}^{ij}(x) \delta^{(3)}(x-y) \,, \vphantom{\ds\int} \\
\pb{\phi(BC\beta)^{a}(x)}{\chi(A)_{b}{}^i(y)} & =  f{^a{}_{bc}}\, \chi(A)^{c}{}{}^i(x)\, \delta^{(3)}(x-y) \,, \vphantom{\ds\int} \\
\pb{\phi(BC\beta)^{a}(x)}{\chi(\beta)_\alpha{}^{jk}(y)} & =  \trg{^{\g a}{}_{\alpha}}\, \chi(\beta)_{\g}{}^{jk}(x)\, \delta^{(3)} (x-y)\,, \vphantom{\ds\int} \\
\pb{\phi(BC\beta)^{a}(x)}{\chi(C)_\alpha{}^i(y)} & =  -\trg{_\alpha{}^{a}{}_\beta}\, \chi(C)^{\beta i}(x)\, \delta^{(3)}(x-y) \,, \vphantom{\ds\int} \\
\pb{\phi(BC\beta)^{a}(x)}{\chi(B)_{b}{}^{jk}(y)} & =  -f{^a{}_{bc}}\, \chi(B){}^{cjk}(x)\, \delta^{(3)}(x-y) \,. \vphantom{\ds\int} 
\end{aligned}
\end{equation}
The elimination of the 2nd class constraints can be achieved by using the Dirac brackets (DB). It can be shown that the DB algebra of the FC constraints is the same as the PB algebra (\ref{AVPK}).

Note that the constraints (\ref{phi}) and the algebra (\ref{AVPK}) reduce respectively to (\ref{cnstr}) and (\ref{algS}), if we consider the second-class constraints (\ref{chi}) as gauge-fixing conditions. 

\section{\label{dof}The physical degrees of freedom} 
  
In this section we will show that the structure of the constraints implies that there are no local degrees of freedom in a $BFCG$ theory. In general case, if there are $N$ initial fields in the theory and there are  $F$ independent first-class constraints per space point and $S$ independent second-class constraints per space point, then the number of local DOF, i.e. the number of independent field components, is given by
\begin{equation} \label{ndof}
n = N - F - \frac{S}{2}\,.
\end{equation}
The formula (\ref{ndof}) is a consequence of the fact that $S$ second-class constraints are equivalent to vanishing of $S/2$ canonical coordinates and $S/2$ of their momenta. The $F$ first-class constraints are equivalent to vanishing of $F$ canonical coordinates, and since the first-class constraints generate the gauge symmetries, we can impose $F$ gauge-fixing conditions for the corresponding $F$ canonical momenta. Consequently there are $2N - 2F - S$ independent canonical coordinates and momenta and therefore $2n = 2N-2F-S$.

In our case, $N$ can be determined from the table 
$$
\begin{array}{|c|c|c|c|} \hline
A^{a}{}_{\mu} & \beta^\alpha{}_{\mu\nu} & C^\alpha{}_{\mu} & B^{a}{}_{\mu\nu} \\ \hline
4p & 6q & 4q & 6p \\ \hline
\end{array}
$$
where $p$ is the dimensionality of the Lie group $G$ and $q$ is the dimensionality of the Lie group $H$. Consequently $N=10(p+q)$.  Similarly, the number of independent components for the second class constraints is determined by the table
$$
\begin{array}{|c|c|c|c|} \hline
\chi(B)_{a}{}^{jk} & \chi(C)_\alpha{}^i & \chi(A)_{a}{}^i & \chi(\beta)_\alpha{}^{ij} \\ \hline
3p & 3q & 3p & 3q \\ \hline
\end{array}
$$
so that $S=6(p+q)$.

The first-class constraints are not all independent, since they satisfy the following relations
\begin{equation}\label{OSIR}
	\begin{aligned}
		\nabla_i \phi(\H)_a{}^{i} + \frac{1}{2}\, \partial_{\alpha a} \,\phi(\G)^\alpha &- \frac{1}{2} \,\partial^\alpha{}_a \, \nabla^\trg_i \chi(C)_\alpha{}^i \\
												&- \frac{1}{2} \,f_{abc}\, \partial_{\alpha}{}^b \,\beta^\alpha_{\,\,\, ij}\, \chi(B)^{c\,ij}=0\,,
\end{aligned}
\end{equation}
\begin{equation}\label{OSIT}
\begin{aligned}
	\ds \nabla^\trg_i \phi(CB)_\alpha{}^{i} &- \frac{1}{2}\,C_{\beta i} \trg{^\beta{}_a{}_\alpha} \phi(\H)^{a i} + \partial_{\alpha a} S(BC\beta)^a + \frac{1}{2}  F^b{}_{ij}\trg_{\alpha b \g} \chi(\beta)^{\g ij}  \\
						&\ds +  T^\beta{}_{jk} \trg_{\beta a \alpha} \chi(B)^{a\,jk} - \partial_{\alpha a} \nabla_i \chi(A)^{a\,i}=0 \vphantom{\ds\frac{1}{2}}\,. \\
\end{aligned}
\end{equation}
One can show that 
\begin{equation}\label{ra}
\begin{aligned}
	\ds  \nabla_i \phi(\H)_a{}^{i} + \frac{1}{2} \partial_{\alpha a} \phi(\G)^\alpha &- \frac{1}{2} \partial^\alpha{}_a  \nabla^\trg_i \chi(C)_\alpha{}^i   \\
											 &\ds  - \frac{1}{2} f_{abc} \partial_{\alpha}{}^b \beta^\alpha{}_{ij} \chi(B)^{c\,ij}  =  \epsilon^{ijk}\nabla_iF_{ajk} \,, \\
\end{aligned}
\end{equation}
which gives (\ref{OSIR}) because $\ep ^{ijk}\, \nabla_i F_{jk}^a =0$ is the $\lambda=0$ component of the BI (\ref{bia}). Similarly
\begin{equation}\label{rb}
\begin{aligned}
	\ds \nabla^\trg_i \phi(CB)_\alpha{}^{i} &- \frac{1}{2}\,C_{\beta i} \trg{^\beta{}_a{}_\alpha} \phi(\H)^{a i} + \partial_{\alpha a} S(BC\beta)^a\\
						&+ \frac{1}{2}  F^b{}_{ij}\trg_{\alpha b \g} \chi(\beta)^{\g ij}  + T^\beta{}_{jk} \trg_{\beta a \alpha} \chi(B)^{a\,jk} \\
						&- \partial_{\alpha a} \nabla_i \chi(A)^{a\,i} = \ep^{ijk}\left( \nabla^\trg_i T_{\alpha jk} -\trg_{\alpha a\beta} F^a_{jk}C^\beta_i \right)\,. \vphantom{\ds\int} \\
\end{aligned}
\end{equation}
The right-hand side of the equation  (\ref{rb}) is the $\lambda=0$ component of the Bianchi identity (\ref{bic}), so that (\ref{rb}) gives the relation (\ref{OSIT}).
   
As discussed in \cite{MOV} for the case of the Poincar\'e 2-group, only the $\lambda=0$ components of the BI give new restrictions on the canonical variables, because those BI do not contain the time derivatives of the fields. The BI components with $\lambda\ne 0$ will contain the time derivatives of the fields and hence must be consequences of the equations of motion (EOM). Related to this is the fact
that the Bianchi identities associated to the 2-forms $\beta$ and $B$ do not induce any new relations among the constraints, see \cite{MOV}. Namely, the corresponding BI (\ref{b2c}) and (\ref{b2d}) contain the time derivatives of the fields, so that the equations (\ref{b2c}) and (\ref{b2d}) are necessarily consequences of the EOM, and hence do not represent additional restrictions on the canonical variables.

The number of components of the first-class constraints can be obtained from the table
$$
\begin{array}{|c|c|c|c|c|c|c|c|} \hline
\phi(B)_{a}{}^i & \phi(C)_\alpha & \phi(\beta)_\alpha{}^i & \phi(A)_{a} & \phi(\H)^{ai} & \phi(\G)^\alpha & \phi(CB)^{\alpha i} & \phi(BC\beta)^{a} \\ \hline
3p & q & 3q & p & 3p & q & 3q & p \\ \hline 
\end{array}
$$
The number of independent components for the first-class constraints is given by
$$F=8(p+q) - p - q = 7(p+q) \,,$$ 
where we have subtracted the $p$ relations (\ref{OSIR}) and the $q$ relations (\ref{OSIT}). Therefore,
\begin{equation} \label{BrojFizickihStepeniSlobode}
n = 10(p+q) - 7(p+q) - \frac{6(p+q)}{2} = 0\,,
\end{equation}
and therefore there are no local DOF in a $BFCG$ theory. Hence the physical DOF are global, and can be identified with the coordinates on the moduli space of the flat 2-connections on the 3-manifold $\Sigma$, see \cite{MO} for the case of the Poincar\'e 2-group. 

\section{\label{conc}Conclusions}

Our canonical analysis implies that the $BFCG$ action (\ref{action}) is a topological field theory action, i.e. an action which is diffeomorphism invariant and has no local DOF. The propagating DOF are global and the corresponding configuration space is the moduli space of flat 2-connections for the $BFCG$ Lie 2-group on the spatial manifold $\Sigma$.

A quantization of a topological field theory can give a topological quantum field theory, which can be then used to obtain manifold invariants, see \cite{W}. In our case, a natural next step would be a canonical quantization of the Hamiltonian formulation of the $BFCG$ theory. For this task it would be important to better understand the PB algebra of the first-class constraints (\ref{AVPK}) and how it is related to the Lie 2-group differential crossed module. 

The $BFCG$ canonical formulation will be also the starting point for constructing the Lie 2-group analogs of holonomy and flux variables from canonical Loop Quantum Gravity (LQG), see \cite{MO}. These variables will be important for the task of finding categorical generalizations of LQG with matter, since the matter is described by the Standard Model Lie group.

\subsection*{Acknowledgments}

AM was partially supported by the FCT projects PEst-OE/MAT/UI0208/ 2013, EXCL/MAT-GEO/0222/2012, and by the bilateral project ``Quantum Gravity and Quantum Integrable Models - 2015-2016'' (451-03-01765/2014-09/24) between Portugal and Serbia. MAO was supported by the FCT grant SFRH/BD/79285/2011.
%MV was supported by the bilateral project ``Quantum Gravity and Quantum Integrable Models - 2015-2016'' (451-03-01765/2014-09/24) between Portugal and Serbia, and by the project ON-171031 of the Ministry of Education, Science and Technological Development, Serbia.
MV was supported by the Science Fund of the Republic of Serbia, grant 7745968, ``Quantum Gravity from Higher Gauge Theory 2021'' --- QGHG-2021. The contents of this publication are the sole responsibility of the authors and can in no way be taken to reflect the views of the Science Fund of the Republic of Serbia.

\providecommand{\href}[2]{#2}

\clearpage

\address{
Departamento de Matem\'atica, Universidade Lus\'ofona\\
Av. do Campo Grande 376, 1749-024 Lisboa, Portugal\\
and Faculdade de Ci\^ encias da Universidade de Lisboa\\
Campo Grande, Edif\'icio C6, 1749-016  Lisboa, Portugal\\
\email{amikovic@ulusofona.pt}}

\address{Faculdade de Ci\^ encias da Universidade de Lisboa\\
Campo Grande, Edif\'icio C6, 1749-016  Lisboa, Portugal\\
\email{masm.oliveira@gmail.com}}

\address{Institute of Physics, University of Belgrade\\
Pregrevica 118, 11080 Belgrade, Serbia\\
\email{vmarko@ipb.ac.rs}}

\end{document}